\documentstyle[12pt,epsfig,amsmath]{article}
\begin{document}


\title{The $s$-$\bar{s}$ asymmetry in nucleon and ``NuTeV
anomaly"}
\author{
F.~X.~Wei~$^{1,2)}$\thanks{weifx@ihep.ac.cn},
B.~S.~Zou$^{1,2,3)}$\thanks{zoubs@ihep.ac.cn}  \\
1) Institute of High Energy Physics, CAS, P.O.Box 918 (4), Beijing 100049\\
2) Graduate University, Chinese Academy of Sciences, Beijing
100049 \\
3) Center of Theoretical Nuclear Physics, National Laboratory of\\
Heavy Ion Accelerator, Lanzhou 730000 }
\date{October 25, 2007}
\maketitle

\begin{abstract}
The $s$-$\bar{s}$ asymmetry in nucleon sea is an important
observable for understanding nucleon structure and strong
interaction. There have been many theoretical attempts on this
subject and recently on its relation to the ``NuTeV anomaly".
Calculations with different theoretical frameworks lead to
different conclusions. Here assuming a newly proposed penta-quark
configuration for the  $s$-$\bar{s}$ asymmetry in nucleon, we
examine its contribution to the ``NuTeV anomaly", with a result of
about $10-20\%$.
\end{abstract}


\section{Introduction}

As a modern theory of strong interaction, quantum chromodynamics
(QCD) is supposed to give us the possibility to describe all
properties of observed hadrons, such as the structure of nucleon.
However, due to its nonperturbative difficulty in infrared region
and the complexity of hadronic phenomena, this is still impossible.
We have to rely on the QCD-inspired phenomenological models, such as
bag model and constituent quark model, to describe effectively some
properties of observed hadrons. With their close relation to the
experimental observables, the deeper investigations to these models
are expected to give some hints for the solutions of QCD, or strong
interaction. The strangeness in nucleon may provide important
observables for studying various models.

According to the quark parton model, which is the consequence of
QCD, a nucleon is composed of 3 valence quarks plus a fluctuating
number of gluons and sea quark anti-quark pairs. Since the strange
quarks are the lightest quarks different from nucleon's valence
quarks, the strangeness in the nucleon is of particular interest
for understanding the role of sea quarks. Experiments have
indicated that strange quarks do, in fact, play a fundamental role
in understanding properties of the nucleon \cite{gas91}. It could
be interpreted that the existence of strangeness in nucleon is a
nonperturbative effect. Then the question can be asked, in what
kind of form do these strange quarks exist in the nucleon ? Many
models have been proposed. The widely used ones are meson cloudy
model and chiral constituent quark model.

Recently, in order to explain the empirical indications for a
positive strangeness magnetic moment of the proton, a new possible
configuration has been proposed for the strangeness in the proton
\cite{zou05}, {\sl i.e.}, the $\bar s$ in the ground state and the
$uuds$ system in the $P$ state. The new configuration can also
reproduce other strangeness properties of the proton
\cite{an206,ris06} and has been successfully extended to explain
properties of other baryons \cite{an06,liqb}. In order to further
check the validity of the new configuration, study of the asymmetry
of parton distribution functions $s(x)$ and $\bar{s}(x)$ versus the
momentum fraction $x$ and its consequence would be a proper choice.
The possible asymmetry of $s(x)$ and $\bar{s}(x)$ has been discussed
in Ref. \cite{sig87} by Signal and Thomas and further explored by
other authors \cite{bro96}. The analysis of related experimental
data \cite{baz95,oln05} seems not conclusive, and the limit of the
$s$-$\bar{s}$ asymmetry quoted in \cite{oln05} is $-0.001 < [S^{-}]
< 0.004$, where $[S^{-}] =\int_{0}^{1} dx x[s(x) - \bar{s}(x)]$. The
refreshed interest on this subject is prompted by the ``NuTeV
anomaly" \cite{kre04} - a 3$\sigma$ deviation of the NuTeV measured
value of $sin^{2}\theta_W$ (0.2277 $\pm$ 0.0013$\pm$0.0009)
\cite{zel02} from the world average of other measurements
(0.2227$\pm$0.0004). The contribution of $s$-$\bar{s}$ asymmetry to
this departure has been discussed in Refs. \cite{kre04,dav02}, and
calculated in Ref. \cite{cao03,alw,din05} in the framework of
meson-baryon model and chiral constituent quark model, respectively.
It is rather puzzling that the two pictures give entirely different
results.

In this article, we will discuss the difference of the meson-baryon
model and chiral constituent quark model related to the strange
content in nucleon; then calculate the second moment of the
strange-antistrange distributions $[S^{-}]$ and its contribution to
the ``NuTeV anomaly" with a newly proposed penta-quark model for the
strangeness in the proton \cite{zou05}.

\section{ The strange parton distribution in nucleon }

The strange quark in nucleon sea, as well as $u$ and $d$, can be
broken down into perturbative and nonperburbative parts. The
perturbative part of $s\bar{s}$ due to short-range fluctuation of
gluon field has no contribution to the $s$-$\bar s$ asymmetry.  We
only focus on the nonperburbative part, which can exist over the
longer time than the interaction time in the deep inelastic process
and hence contributes to the $s$-$\bar s$ asymmetry observables. As
discussed in precious section, there are many models about the
nonperturbative strange sea quarks. These models can be classified
into two sorts: meson-baryon configuration and quark-meson
configuration. The dynamical information of the two pictures can be
obtained from relevant scattering experiments.

In the meson-baryon configuration, the nucleon sometimes fluctuates
to a baryon plus a meson. Contributions to the strange sea can come
from fluctuations involving a hyperon, such as
$p(uud)\to\Lambda(uds) + K^{+}(u\bar{s})$. In this example, the
contribution to the strange quark distribution s(x) comes from the
strange quark in the $\Lambda$, while the contribution to the
anti-strange distribution $\bar{s}(x)$ comes from the anti-strange
quark in the kaon. Then the strange distribution can be calculated
by using the valence parton distribution of $\Lambda$ and kaon,
respectively. Because of the different fluctuation functions and
different parton distributions in $\Lambda$ and $K^{+}$, the
calculated results for $s(x)$ and $\bar{s}(x)$ are different.
However, there are some theoretical uncertainties in this picture.
First, the dynamical quantities, such as coupling constants, which
are derived from reproducing experimental data on scattering
processes, may be invalid in applying directly to the interior of
the nucleon. The off-shell extension suffers large uncertainty.
Secondly, the parton distributions of $\Lambda$ and $K^{+}$, which
can not be directly calculated from first principle, would also
bring large uncertainties to the results. We will see that these two
problems could be avoided in the penta-quark model.

In the chiral constituent quark model with quark-meson
configuration, the meson octet was introduced as the Goldstone
particles, which are the consequences of the spontaneously broken
chiral symmetry (SBCS). Therefore, the quarks are dressed by mesons.
The relevant degrees of freedom in this configuration are
constituent quarks and Goldstone bosons (The effect of gluon can be
negligible at low energy). In this picture, the constituent quarks
couple directly to the GS bosons, for example, $u \to K^+(u\bar{s})
+ s $. The contribution to $\bar s(x)$ comes from the parton
distribution in $K^{+}$, and contribution to $s(x)$ comes directly
from dynamical process. Obviously, this picture also results in
different $s(x)$ and $\bar{s}(x)$ distributions. Because the SBCS is
included in this configuration, which is the nonperturbative effect
of QCD, this picture is expected to provide a satisfactory
representation for low energy hadron properties.

The results of the $s(x)$ and $\bar{s}(x)$ in these two
configurations are very different, and even contradictory in some
special regions \cite{cao03,din05}. And the predicted $s$-$\bar{s}$
asymmetry from these two pictures differs by about two orders of
magnitude. While the calculation within the framework of effective
chiral quark model claims that the $s$-$\bar{s}$ asymmetry can
account for about $60-100\%$ of the NuTeV anomaly \cite{din05}, the
calculations with the meson-baryon configuration give much smaller
results ranging from $1\%$ \cite{cao03} to $20\%$ \cite{alw}.
Besides the choice of parameters, the interaction of $s$ and
$\bar{s}$ with other constituents may be the key to understand this
difference. In meson-baryon configuration, the $s$ is bound in the
hyperon, while the $s$ is asymptotic free in the meson-quark
picture. We reckon that the fluctuations of $q(u,d) \to Ks$ give a
harder momentum distribution for $s$ than that given by nucleon
fluctuations into $|BM \rangle$.

Recently a new possible configuration for the five quark components
in the nucleon has been proposed \cite{zou05,an206,ris06,an06}. In
the penta-quark model, the largest five quark components in the
proton are $uudd\bar d$ and $uuds\bar s$ with the anti-quark in the
orbital ground state and the four quarks in the mixed orbital
$[31]_X$ symmetry, i.e., one in P-wave and three in S-wave.
Therefore, the quark wave function for the proton may then be
expanded as:
\begin{equation}
|p> = A_{3q}|uud> + A_{d\bar d} |[ud][ud]\bar d> + A_{s\bar s}
|[ud][us]\bar s>\,
\end{equation}
with the normalization condition $|A_{3q}|^2+|A_{d\bar d}|^2+
|A^2_{s\bar s}|^2=1$. The fluctuation probability of the $d\bar{d}$
and $s\bar{s}$, which are interpreted as the probability to find the
$uudd\bar{d}$ component and $uuds\bar{s}$ in a proton, can be
obtained as $P_{d\bar d}\equiv |A_{d\bar d}|^2 =12\%$ and $P_{s\bar
s}\equiv |A_{s\bar s}|^2 = (12-48)\%$ by reproducing the observed
light flavor sea quark asymmetry in the proton, $\bar d-\bar u=0.12$
and the strangeness spin of the proton, $\Delta_s=-0.10\pm 0.06$ ,
respectively.

Since the $\bar{s}$ in $uuds\bar{s}$ system is in its ground state
and the $uuds$ subsystem has mixed orbital symmetry $[31]_{X}$ which
gives the possibility of 1/4 for $s$ to be in P-wave, this also
gives naturally an $s$-$\bar s$ asymmetry.

\section{ NuTeV anomaly and contribution from $s$-$\bar{s}$ asymmetry}

The ''NuTeV anomaly" is an important open question in recent years.
Although many sources of it have been explored in the past years
\cite{kre04,dav02} there has been no consistent explanation on this
subject. The measurement of Weinberg angle $\theta_{W}$ in Ref.
\cite{zel02} by NuTeV collaboration is closely related to the
Paschos-Wolfestein(PW) relation \cite{pas73}, which is written as
\begin{equation}
R^{-} \equiv \frac{\sigma^{\nu N}_{NC} -
\sigma^{\bar{\nu}N}_{NC}}{\sigma^{\nu N}_{CC} -
\sigma^{\bar{\nu}N}_{CC}} \simeq  1/2 - sin^{2}\theta_W + \delta
R^{-}_{A} + \delta R^{-}_{QCD} + \delta R^{-}_{EW},
\end{equation}
where the three $\delta$ terms are due to the nonisoscalarity of the
target ($\delta R^{-}_{A}$),next-to-leading-order(NLO) and
nonperturbative QCD effects ($\delta R^{-}_{QCD}$), and higher-order
electroweak effects ($\delta R^{-}_{EW}$), respectively. The QCD
corrections consist of three terms, which can be written as $\delta
R^{-}_{QCD} = \delta R^{-}_{s} + \delta R^{-}_{I} + \delta
R^{-}_{NLO}$, where the three $\delta$ terms in the right side are
due to possible strange asymmetry ($ \delta R^{-}_{s}$) and isospin
violation ($u_{p,n} \neq d_{n,p}$) effects ($\delta R^{-}$) in the
parton structure of nucleon, and NLO($O(\alpha_{s})$)
corrections($\delta R^{-}_{NLO}$), respectively. In this paper, we
only focus on the correction from $s$-$\bar{s}$ asymmetry, which
contributes to $R^{-} $ as
\begin{equation}
\delta R^{-}_{s} \simeq -(\frac{1}{2} -
\frac{7}{6}sin^{2}\theta_W)\frac{[S^{-}]}{[Q^{-}]},
\end{equation}
where $[S^{-}] \equiv \int x[s(x) - \bar{s}(x)]dx$ quantifying the
strangeness asymmetry, and $[Q^{-}] = \int x[q(x) - \bar{q}(x)]dx$
with $q(x) = [u(x) + d(x)]/2$ representing the isoscalar valence
quark distribution. In order to solve the NuTeV anomaly, the sigh of
$[S^{-}]$ needs to be positive, i.e. $[S^{-}]
> 0$.

Generally, for a nucleon in its $|A,B\rangle$ configuration created
in the fluctuation process $|N \rangle \to |A\rangle + |B\rangle $
with $s$ and $\bar{s}$ in $|A \rangle$ and $|B \rangle$,
respectively, the $s$ distribution can be expressed as a convolution
of fluctuation function $f_{AB/N}(x)$ with the valence parton
distribution $s_{A}(x)$ in the state $|A \rangle$; and the
distribution of $\bar{s}$ can be expressed as a convolution of
fluctuation function $f_{BA/N}(x)$ with the valence parton
distribution $\bar{s}_{B}(x)$ in the state $|B \rangle$
\cite{hol96}. Explicitly, the strange and anti-strange quark
distributions in the nucleon can be written as
\begin{eqnarray}
s(x) = \int_{x}^{1} \frac{dy}{y} f_{AB/N}(y) s_{A}(\frac{x}{y}),\\
\bar{s}(x) = \int_{x}^{1} \frac{dy}{y} f_{BA/N}(y)
\bar{s}_{B}(\frac{x}{y}) .
\end{eqnarray}
with general constraints $f_{AB/N}(x) = f_{BA/N}(1-x)$ and
$\int_{0}^{1} dx f_{BA/N}(x) = \int_{0}^{1} dx f_{BA/N}(x) =
P_{AB/N}$, where $P_{AB/N}$ is the probability to find the $\mid A,B
\rangle$ configuration in a nucleon.

The fluctuation function $f_{AB/N}(x)$ is interpreted as probability
to find $|A\rangle$ with a fraction $x$ of the nucleon momentum,
while the $f_{BA/N}(x)$ is the probability to find $|B\rangle$ with
a fraction $x$ of nucleon momentum. It reflects the dynamical
information of the fluctuation process, which is the nonperturbative
effect closely related to QCD at large distances. The dynamical
mechanism behind this process may be important for further research.

However, in our case with penta-quark configuration, the thing is
getting simpler. The fluctuation function is just $f_{5q/N}(x) =
P_{5q/N}\delta(x-1)$. The dynamical information of the fluctuation
process is included into the probability which can be obtained from
experimental data. This could be one of advantages of the
penta-quark model.

The next step in our calculation is to determine parton distribution
in the penta-quark configuration. Simple harmonic oscillator wave
functions are used with radial part as
\begin{eqnarray}
\varphi^{S}(k)&=&{1\over (\alpha^2\pi)^{3/4}}
\exp(-\frac{k^{2}}{2\alpha^{2}}),\\
\varphi^{P}(k)&=&{k\over \alpha}\,\varphi^{s}(k)\,,
\end{eqnarray}
for the S-state and P-state, respectively. Here $\alpha^{2} =
m_{s}\omega$ with $\omega$ the harmonic oscillator parameter. With
these wave functions, the distributions of $s$ and $\bar{s}$ in
$uuds\bar{s}$ system can be obtained by the method in Ref.
\cite{sig89,sch91}, in which the distributions of $s(\bar{s})$ in
the five-quark constituent can be expressed as
\begin{eqnarray}
s_{5q}(x) &=& \int d\vec{k} \delta(Mx - k^{+}) (\frac{3}{4}
|\varphi^S(k)|^{2} + \frac{1}{4}| \varphi^P(k)|^{2}) , \\
\bar{s}_{5q}(x) &=& \int d\vec{k} \delta(Mx - k^{+})
|\varphi^S(k)|^{2} ,
\end{eqnarray}
where M is the mass of the nucleon and $k^{+}$ the light-cone
momentum of $s(\bar{s})$. The $s_{5q}(x)$ and $\bar{s}_{5q}(x)$ need
to be normalized to 1.

\begin{figure}[h]
\label {fig1} \centering
\mbox{\psfig{figure=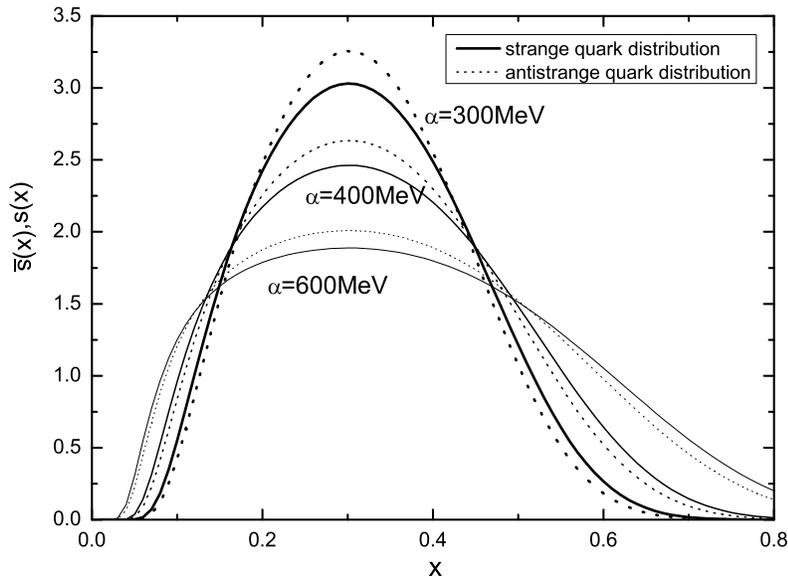,height=0.4\textheight, clip=}}
\caption{The distributions of strange quarks (solid), $s(x)$, and
antistrange quarks (dotted), $\bar{s}(x)$, in $uuds\bar{s}$ system.}
\end{figure}

Assuming commonly used values $m_{s} = m_{\bar{s}} = 400MeV$ and
$\alpha = 300, 400, 600MeV$, the calculated results of $s_{5q}(x)$
and $\bar{s}_{5q}(x)$ are shown in Fig.1.  Compared with $\bar{s}$,
the $s(x)$ is softer in small x and harder in large x region. This
variation can be easily understood in our theoretical frame, because
in the penta-quark model the difference of $s(x)$ and $\bar{s}(x)$
in nucleon entirely results from the different distributions of $s$
and $\bar{s}$ in the $uuds\bar{s}$ component. While the $\bar s$
stays 100\% in S-state, the $s$ has 25\% probability staying in
P-state. Hence the $s$ is more likely to take larger fraction of
nucleon momentum.  The distribution of $x\delta_{s}(x)$, with
$\delta_{s}(x) = s(x) - \bar{s}(x)$ is shown in Fig. 2. The behavior
of $x \delta s(x)$ can be well understood in the penta-quark
configuration where the $\bar{s}$ stays in the S-wave around the
center of the system while $s$ in the $uuds$ has 25\% probability in
the P-state and gives harder distributions ($s(x)$) at large $x$
region.

\begin{figure}[h]
\label {fig2} \centering
\mbox{\psfig{figure=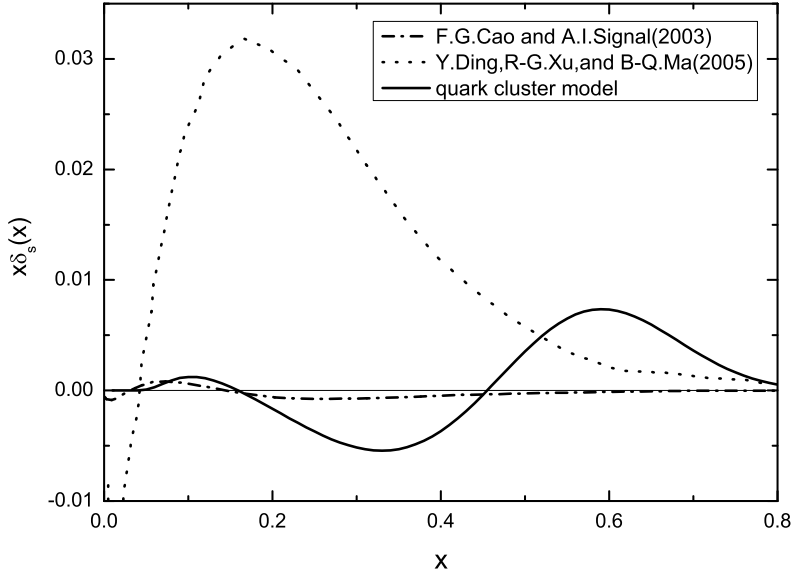,height=0.4 \textheight, clip=}}
\caption{The distribution of $x\delta_{s}(x)$, with $\delta_{s}(x) =
s(x) - \bar{s}(x)$.}
\end{figure}

The results are sensitive to the value of parameter $\alpha$. Larger
$\alpha$ leads to larger difference of $s(x)$ and $\bar{s}(x)$. This
is because larger $\alpha$ gives larger difference between P-state
and S-state. The choice of $m_{s}$ value makes little effect on the
result.

>From these strange parton distributions, assuming $\alpha = 400MeV$
and $P_{s\bar{s}} = 20\%$, we obtain $[S^{-}] = 0.001$, which can
account for 10\% of the NuTeV anomaly. There is some evidence
\cite{an06} suggesting that the $qqqs\bar{s}$ constituent is very
compact with $\alpha$ around $1 GeV$. In this case, the
$s$-$\bar{s}$ asymmetry would result in $[S^{-}] = 0.002$ and
account for about $20\%$ of NuTeV anomaly. The result from the newly
proposed penta-quark configuration is much smaller than that from
chiral constituent quark model \cite{din05}, but comparable with
that of meson-baryon models \cite{cao03,alw}. As shown in Ref.
\cite{kre04,dav02}, there are many other uncertainties in
theoretical framework for NuTeV experiment, and some other
corrections may account for the NuTeV anomaly. The $s$-$\bar{s}$
asymmetry may not be the whole story.

\bigskip
\noindent {\bf Acknowledgement:} We would like to thank B.Q.Ma for
useful discussions and J.J.Wu for a help on integration program.
This work is partly supported by the National Natural Science
Foundation of China under grants Nos. 10435080, 10521003 and by the
Chinese Academy of Sciences under project No. KJCX3-SYW-N2.

\end{document}